\documentclass[11pt]{article}
\usepackage{graphicx}

\usepackage{amsmath}
\usepackage{color}
\usepackage{bbm}
\usepackage{ulem}

\newcommand{\cC}{{\cal C}} 
\newcommand{\cF}{{\cal F}} 
\newcommand{\cI}{{\cal I}}

\newcommand{\bbF}{{\bf F} }

\newcommand{\bra}{\langle}
\newcommand{\ket}{\rangle}

\newtheorem{rem}{Remark}[section] 

\newcommand{\cond}{\Sigma}

\title{Phase fluctuations in  the ABC model}
\date{\today}
\author{T. Bodineau\footnote{Ecole Normale Sup\'erieure, DMA, 45 rue d'Ulm
75230 Paris cedex 05, France},
B. Derrida\footnote{Laboratoire de Physique Statistique
(CNRS UMR 8550), \'Ecole Normale Sup\'erieure,
UPMC Paris 6, Universit\'e Paris Diderot Paris 7, CNRS,
 24 rue Lhomond, 75231 Paris cedex 05, France}
\\
}

\begin{document}

\maketitle

\begin{abstract}
We analyze the fluctuations of the steady state profiles in the modulated phase of the ABC model. 
For a system of $L$ sites, the steady state profiles move on a microscopic time scale of order $L^3$. 
The variance of their displacement  is computed in terms of the macroscopic steady state profiles by using fluctuating hydrodynamics and large deviations. 
Our analytical prediction for this variance is confirmed by the results of numerical simulations.
\end{abstract}


\section{Introduction}

A surprising property of   non-equilibrium systems in their steady state  is that they may exhibit  phase transitions  in one dimension
\cite{Krug,DDM,Mallick,Evans,schutz1,KLMST, PFF, EFGM,GSW}.
One of the  simplest models for  which  such a phase transition occurs is the ABC model \cite{EKKM1,EKKM2,CDE,FF1,FF2,ACLMDS,LM,BLS1,LCM}.
 In the ring geometry  \cite{EKKM1,EKKM2,CDE,FF1,FF2} the model describes, on a ring of $L$ sites,  a  lattice gas of three species of particles  undergoing  asymmetric exchanges  between neighboring sites: each lattice site is occupied  by a particle  of type  either A,  or B or C and particles on neighboring sites exchange with the following rates

\begin{eqnarray} AB & {{q \atop \longrightarrow}\atop
{\longleftarrow \atop 1}} & BA
\nonumber \\ BC & {{q \atop \longrightarrow}\atop
{\longleftarrow \atop 1}} & CB
\label{update}
\\ CA & {{q \atop \longrightarrow}\atop
{\longleftarrow \atop 1}} & AC
\nonumber
\end{eqnarray}
where $q \leq 1$.
The total numbers $N_A,N_B,N_C$ (with $N_A+N_B+N_C=1$)  of particles of each species are conserved by the dynamics.
Under these dynamics, the system reaches in the long time limit a steady state.
When $q=1$, all configurations are equally likely in the steady state. Therefore the steady state profiles are flat.
As soon as $q \neq 1$, the translational symmetry is broken and the steady state consists of three macroscopic domains corresponding to the three species \cite{EKKM1,EKKM2}.
The phase transition    between these two regimes for the ring geometry has been studied in \cite{CDE,FF1,FF2,CM}. More recent works have also considered the ABC model on an open interval \cite{ACLMDS} and in its grand canonical version \cite{LM,BLS1,LCM}.

When the asymmetry $q$ in (\ref{update})  scales as
\begin{equation}
q =\exp \left[ - {\beta \over L} \right]  
\label{q-scaling}
\end{equation}
the dynamics is diffusive. This means that, on time scales of order $L^2$,  one can describe the system by density profiles $\{\rho_A(x,\tau),\rho_B(x,\tau),\rho_C(x,\tau)\}$  functions of a macroscopic  coordinate $0<x<1$ and of  a macroscopic time  $\tau$ which are related to the labels $k$  of the lattice sites ($1 \leq k \leq L$)  and to the microscopic time $t$ by a diffusive scaling
\begin{equation}
\label{scaling}
k= L x \ \ \ \ \  \ ; \ \ \ \  t=L^2 \tau
\end{equation}

In the ring geometry
 when   one varies the asymmetry  $\beta$ in (\ref{q-scaling}), one observes a phase transition  \cite{EKKM1,EKKM2,CDE,FF1,FF2} in the steady state, from a flat phase where the density profiles  $\{ \rho_A(x),\rho_B(x),\rho_C(x) \} $
of the three species    do not depend on the position $x$ on the ring  to   a segregated phase where  the profiles  become modulated and therefore space dependent (on the ring  
 all the  density profiles are periodic  functions of  the macroscopic  coordinate $x$, with period $1$).
As the dynamical rules in the ABC model are translation invariant,   
 the  continuous translational symmetry is broken in the segregated phase. Therefore, in the steady state,  if one observes a set of three density profiles  $\{ {\bar \rho_A}(x),{\bar \rho_B}(x), \bar\rho_C(x) \}$,       the same density profiles 
\begin{equation}
\{ \bar\rho_A(x -\xi), \bar\rho_B(x -\xi), \bar\rho_C(x - \xi) \}
\label{profiles}
\end{equation}
translated by an arbitrary amount $\xi$ along the circle are also  steady state profiles.
For general densities of the three species, one does not know 
if the shape of the steady state profiles is unique. 
Throughout this paper, we will however assume that the uniqueness holds.

In the thermodynamic limit,  each set of the profiles (\ref{profiles}) labelled by a  fixed value of $\xi$     are steady state   profiles, in the modulated phase,  with an  infinite life time and ergodicity is broken.
On the other hand, for a large but finite system size $L$, the steady state is unique and    the density profiles get translated by   a time dependent  random amount $\xi_\tau$ due to   the stochastic dynamics. Thus    one observes in the steady state the following time dependence  of  the three profiles
\begin{equation}
\label{eq: 4 bis}
\{ \bar \rho_A(x-\xi_\tau), \bar \rho_B(x-\xi_\tau), \bar\rho_C(x-\xi_\tau) \}
\end{equation}
where the phase $\xi_\tau$ performs a stochastic motion along the ring.
The goal of the present paper is to   characterize the fluctuations of the phase $\xi_\tau$  
and to show that for a large system size its variance is given  by
\begin{equation}
\lim_{\tau \to \infty} \frac{1}{\tau} \big( \langle \xi_\tau^2 \rangle - \langle \xi_\tau \rangle^2 \big)  \simeq
{D \over L}
\label{main-result}
\end{equation}
where  $L$ is the number of sites on the ring and the diffusion constant $D$ is given in terms of the steady state  density profiles 
\eqref{profiles} by
\begin{equation}
D=
{ 2  \int dx  \left(  \bar \rho_A   \bar\rho_B \bar\rho_C'^2  +    \bar\rho_B  \bar\rho_C  \bar\rho_A'^2 +   \bar\rho_C  \bar\rho_A  \bar\rho_B'^2 \right) \over 
\left[\int dx  \left( \bar \rho_A  \bar\rho_B' -  \bar\rho_A'  \bar\rho_B\right) \right]^{2}}
\label{main-result-ter}
\end{equation}
(One can easily check  that  this last expression is  symmetric in $\bar\rho_A,\bar\rho_B,\bar \rho_C$ using   the fact that 
$\bar\rho_A+\bar\rho_B+\bar \rho_C=1 $ which implies that $\bar \rho_A  \bar\rho_B' -  \bar\rho_A'  \bar\rho_B= \bar \rho_B  \bar\rho_C' -  \bar\rho_B'  \bar\rho_C  + \bar\rho_B'$
so that the squared integral in the denominator of (\ref{main-result-ter}) remains unchanged in any  permutation of the three species.)

The scaling of the variance \eqref{main-result} predicts a microscopic relaxation on a time  $t\sim L^3$  which is  exactly the  same time scale found recently for this problem when the three species have equal densities \cite{BCP}.
Front fluctuations have been already analyzed (in full mathematical rigor) in \cite{BDP,BBDP, BB} for one-dimensional non-conservative dynamics and in \cite{BBBP} for a system of coupled equations.  The relation between front fluctuations and large deviations of forced interfaces has also been studied in \cite{DDP}.

\medskip

The outline of the paper is as follows. In section 2, we briefly recall some known properties of the ABC model on a ring in the diffusive limit.
In section 3, we present the calculation of the variance (\ref{main-result}, \ref{main-result-ter}) using fluctuating hydrodynamics.
In section 4, we give an alternative  derivation of  the diffusion constant  (\ref{main-result-ter}) based on large deviations in the spirit of  \cite{DDP}.
In section 5, we show how our main result  (\ref{main-result}, \ref{main-result-ter}) gets simplified in  the equal density case (where the dynamics satisfies detailed balance) or when one approaches a second order phase transition. In section 6, we present the results of numerical simulations which agree well with the prediction (\ref{main-result}, \ref{main-result-ter}).

\section{The ABC model on a ring}

For the ABC model, on a ring of $L$ sites,  each site $i$ is occupied by one particle of type A, B or C and the exchange rates 
between neighboring sites are given  by
(\ref{update}).  As the dynamics  is ergodic and  conserves  the total number of particles of each species, the steady state  properties depend only on  the total densities $r_A, r_B, r_C$ of the three species which satisfy of course 
\begin{equation}
r_A+ r_B+ r_C =1
\label{relation}
\end{equation}
and on the asymmetry $q$ in the exchange rates (\ref{update}).

When this asymmetry $q$  scales with  the system size  $L$  as  in (\ref{q-scaling}), 
the dynamics  become diffusive and the density profiles $\rho_A(x,\tau), \rho_B(x,\tau),\rho_C(x,\tau)$ evolve in the  infinite  $L$ limit   according to 
 \cite{CDE} 
\begin{eqnarray}
\partial_\tau \rho_A(x,\tau) =  \partial_x^2 \rho_A (x,\tau)  + \beta \partial_x \big( \rho_A (x,\tau)(\rho_B (x,\tau)- \rho_C(x,\tau)) \big)
\nonumber 
\\
\label{eq: hydro}
\partial_\tau \rho_B(x,\tau) =  \partial_x^2 \rho_B (x,\tau)  + \beta \partial_x \big( \rho_B (x,\tau)(\rho_C (x,\tau)- \rho_A(x,\tau)) \big)
\\
\partial_\tau \rho_C(x,\tau) =  \partial_x^2 \rho_C (x,\tau)  + \beta \partial_x \big( \rho_C (x,\tau)(\rho_A (x,\tau)- \rho_B(x,\tau)) \big)
\nonumber
\end{eqnarray}
on macroscospic scales (\ref{scaling}).
To describe the profiles on time scales much larger than $L^2$, one would need to take into account stochastic corrections of order $1/\sqrt{L}$ to these equations (see section \ref{sec: fluct hydro}) and possibly higher order deterministic corrections in $1/L$.

Flat profiles
 $\{\rho_A(x),\rho_B(x),\rho_C(x) \} 
 =\{r_A,r_B,r_C\} $
 are  always steady state  solutions of these equations. It is  however  easy  to see  \cite{CDE} that they are linearly unstable   when  the asymmetry parameter  $\beta$ in (\ref{q-scaling})  exceeds  a  certain critical value
 $\beta_c$ given by
\begin{equation}
\beta_c=
 \frac{ 2\pi}{\left[ 1 - 2 (r_A^2 + r_B^2 + r_C^2)\right]^{1/2}}
\label{betac-unequal}
\end{equation}
So when $\beta> \beta_c$ the steady state profiles are always modulated.

In the equal density case ($r_A=r_B=r_C=1/3$), it is known that  (\ref{betac-unequal}) gives the location 
$ \beta_c= 2\pi \sqrt{3} $
of the phase transition.
 For other densities, (\ref{betac-unequal}) is simply the value  of $\beta$  where flat profiles  become  linearly unstable, and so it coincides with the location  of  the phase transition
between the flat phase and the modulated phase  only  when  this phase transition is second order. One expects  \cite{CDE} in particular that a first order transition should occur  at least when
$r_A^2+r_B^2+ r_C^2 < 2 (r_A^3 + r_B^3 + r_C^3)$ in which case (\ref{betac-unequal}) is certainly not the location of the phase transition.

The reason why the  equal density case
$r_A=r_B=r_C=1/3$  is best understood is that its dynamics satisfies detailed balance  \cite{EKKM1,EKKM2}  and the steady state measure is known.
As a result the steady state profiles in the modulated phase do not move.
For unequal densities $\{r_A,r_B,r_C\} \neq \{1/3,1/3,1/3 \}$, the dynamics does not satisfy detailed balance and the steady state 
could move at a fixed velocity or even have a more complicated time behavior. 
We argue in Appendix I that the velocity should remain  zero for some range of parameters in a neighborhood of the 
equal density case. Throughout this paper, we will assume that the steady state profiles do not move in which case they  satisfy
\begin{eqnarray}
0 =  \partial_x^2 \bar\rho_A   + \beta \partial_x \big( \bar\rho_A (\bar\rho_B - \bar\rho_C) \big)
\nonumber \\
\label{eq: hydro-bis}
0=  \partial_x^2 \bar\rho_B   + \beta \partial_x \big( \bar\rho_B (\bar\rho_C - \bar\rho_A) \big)
\\
0=  \partial_x^2 \bar\rho_C   + \beta \partial_x \big( \bar\rho_C (\bar\rho_A - \bar\rho_B) \big)
\nonumber
\end{eqnarray}

Expressions of these steady state profiles in terms of elliptic integrals have been given in \cite{CM,FF1,ACLMDS}.
As each site is occupied by a particle $A, B$ or $C$, one always has 
\begin{equation}
\rho_A(x,\tau)+
\rho_B(x,\tau)+
\rho_C(x,\tau) =1 
\label{conservation}
\end{equation}
Therefore one can restrict the analysis to the study of only two density profiles with
(\ref{eq: hydro}) replaced by 
\begin{eqnarray}
\partial_\tau \rho_A =  \partial_x^2 \rho_A   + \beta \partial_x \big( \rho_A (\rho_A +2  \rho_B-1) \big)
\nonumber
\\
\label{eq: hydro1}
\partial_\tau \rho_B =  \partial_x^2 \rho_B   + \beta \partial_x \big( \rho_B (1-  2 \rho_A - \rho_B) \big)
\end{eqnarray}
and (\ref{eq: hydro-bis})  by
\begin{eqnarray}
0 =  \partial_x^2 \bar\rho_A   + \beta \partial_x \big( \bar\rho_A ( \bar\rho_A + 2 \bar\rho_B-1) \big)
\nonumber \\
\label{eq: hydro1-bis}
0=  \partial_x^2 \bar\rho_B   + \beta \partial_x \big( \bar\rho_B (1-2  \bar\rho_A - \bar\rho_B) \big)
\end{eqnarray}

 When the system is large but finite, due to the stochastic nature of the microscopic dynamics, the    density profiles $\rho_A(x,\tau), \rho_B(x,\tau)$ (and $\rho_C(x,\tau)$  related to them by (\ref{conservation}))
have no longer a deterministic evolution given by (\ref{eq: hydro}).
This deterministic evolution  is replaced as in the macroscopic fluctuation theory \cite{BDGJL, derrida2007, Spohn}
by a probability distribution of observing 
macroscopic density profiles $\{\rho_A(x,\tau), \rho_B(x,\tau)\}$ during a macroscopic  time interval $\tau_1 < \tau < \tau_2$  given by
\cite{BDLV}
\begin{equation}
 {\rm Pro}(\{\rho_A,\rho_B \}) \sim  \exp \left( - L  \min_{j_A(x,\tau),j_B(x,\tau)}  \ {\cal I}_{[\tau_1,\tau_2]} (\rho_A,\rho_B,j_A,j_B)  \right) 
\label{pro}
\end{equation}
where the large deviation functional is given by
 \begin{eqnarray}
\label{eq: I ABC}
& &    {\cal I}_{[\tau_1,\tau_2]} (\rho_A,\rho_B,j_A,j_B) = 
\\ \nonumber
 & &  \ \ \ \ \ \int_{\tau_1}^{\tau_2} d \tau   
\int_0^1 dx
 {\sigma_{BB} (j_A - q_A)^2 - 2 \sigma_{AB}(j_A- q_A)(j_B-q_B) +
\sigma_{AA} (j_B - q_B)^2 \over 2 ( \sigma_{AA} \sigma_{BB} - \sigma_{AB}^2)} 
\end{eqnarray}
with 
\begin{eqnarray}
&& \sigma_{AA} = \sigma_{AA} (\rho)= 2 \rho_A (1- \rho_A), \qquad
\sigma_{AB}= \sigma_{AB} (\rho)  = - 2 \rho_A  \rho_B  \nonumber \\ 
&& \sigma_{BB} = \sigma_{BB} (\rho) =  2   \rho_B (1 - \rho_B) \label{sigma}
\end{eqnarray} 
and  $q_A,q_B$  given by
\begin{eqnarray}
 \label{qaqb}
 q_A= -{d \rho_A \over dx} - \beta \rho_A(\rho_A+ 2 \rho_B- 1), \quad   
 q_B= -{d \rho_B \over dx} - \beta \rho_B(1- 2\rho_A-\rho_B)
\end{eqnarray}
In (\ref{pro})
the optimum 
 is over the currents $j_A(x,\tau),j_B(x,\tau)$
which satisfy the conservation laws
\begin{eqnarray}
{d \rho_A \over d \tau}= - {d j_A \over dx}, \qquad 
 {d \rho_B \over d \tau}= - {d j_B \over dx} 
\label{conservations}
\end{eqnarray}

An alternative way of representing the noise produced at the macroscopic scale 
by the stochastic dynamics of the microscopic model is to   describe the evolution of the macroscopic profiles by fluctuating hydrodynamics \cite{BDLV}:
 the current profiles are then given by 
\begin{eqnarray}
\label{current}
j_A = q_A + { 1 \over \sqrt{L}} \eta_A(x,\tau), \qquad 
j_B = q_B + { 1 \over \sqrt{L}} \eta_B(x,\tau)
\end{eqnarray}
which together with (\ref{conservations}) give the stochastic evolution of the macroscopic profiles.
In (\ref{current}),   $\eta_A, \eta_B$ are Gaussian  noises, white  in time  and in space, with the following correlations
\begin{eqnarray}
 & & \langle  \eta_A(x,\tau) \rangle =  
  \langle  \eta_B(x,\tau) \rangle = 0 
\nonumber \\
 & & \langle  \eta_A(x,\tau) \eta_A(x',\tau')\rangle =  \sigma_{AA} \big( \rho (x,\tau) \big)   \;  \delta(x-x') \  \delta(\tau-\tau') 
\nonumber \\ 
 & & \langle  \eta_A(x,\tau) \eta_B(x',\tau')\rangle =  \sigma_{AB} \big( \rho (x,\tau) \big) \;  \delta(x-x')  \ \delta(\tau-\tau') 
\label{noise}
\\ \nonumber 
 & & \langle  \eta_B(x,\tau) \eta_B(x',\tau')\rangle =  \sigma_{BB}\big( \rho (x,\tau) \big)  \;  \delta(x-x')  \ \delta(\tau-\tau') 
\end{eqnarray}
where the expressions of $\sigma_{AA}, \sigma_{AB},\sigma_{BB}$ in terms of the time dependent profiles
$\rho_A(x,\tau),\rho_B(x,\tau)$  are given in (\ref{sigma}).

\section{The fluctuating hydrodynamic approach}
\label{sec: fluct hydro}

For large $L$, in the steady state, the probability that the density profiles differ noticeably from the profiles $\bar{\rho}$ is exponentially small.
The main idea  to derive (\ref{main-result}, \ref{main-result-ter}) 
from  the fluctuating  hydrodynamic approach 
is   to write for large $L$ the density profiles as
\begin{eqnarray}
\rho_A(x,\tau)\simeq \bar{\rho}_A(x- \xi_\tau) + {1 \over \sqrt{L}} \phi_A(x - \xi_\tau,\tau)
\nonumber 
\\
\rho_B(x,\tau)\simeq \bar{\rho}_B(x- \xi_\tau) + {1 \over \sqrt{L}} \phi_B(x - \xi_\tau,\tau)
\label{noisy-profiles}
\end{eqnarray}
where  $\bar{\rho}_A$  and $\bar \rho_B$ are the steady state profiles   of the infinite system,   
$\phi_A$ and $\phi_B$ are  the  fluctuating  parts  of these  profiles due to the noise
 and $\xi_\tau$ is the cumulative translation of the profiles due to the noise.
  $\phi_A$ and $\phi_B$ do not grow with time.
The approach we then follow is inspired by the works  \cite{BDP, BBBP,RCEV,Panja,MSK} on the effect of noise on the position of fronts.

In order to manipulate conveniently these quantities, it is  useful to introduce the vectors
\begin{equation}
 \bar\rho(x)  = \left(\begin{array}{c}
\bar\rho_A(x) \\
 \bar\rho_B(x) 
\end{array} \right)
\ \ ; \ \ 
\phi(x,\tau) = \left(\begin{array}{c}
\phi_A(x,\tau) \\
\phi_B(x,\tau) 
\end{array} \right)
\ \ ; \ \ 
\eta(x,\tau) = \left(\begin{array}{c}
\eta_A(x,\tau) \\
\eta_B(x,\tau) 
\end{array} \right)
\end{equation}
and the covariance matrix of the noise (\ref{sigma}, \ref{noise})
\begin{equation}
\Sigma(\bar\rho(x)) = 
\left(\begin{array}{cc}  
\sigma_{AA} (\bar\rho (x) )  &  \sigma_{AB} (\bar\rho (x) ) \\ 
\sigma_{AB} (\bar\rho (x) ) &  \sigma_{BB} (\bar\rho (x) )
\end{array} \right) 
\label{Sigma}
\end{equation}
Let us also define 
 the operator ${\cal L}$, acting on such space dependent  vectors,  
which is obtained by linearizing the macroscopic equations (\ref{eq: hydro1})
around the steady state profiles $\bar \rho$
\begin{equation}
{\cal L}  \left(\begin{array}{l}
\psi_A(x) \\
\psi_B(x) 
\end{array} \right)
= \left(\begin{array}{c}
\partial_x^2\psi_A(x) + \beta \partial_x \Big(  (2 \bar \rho_A(x) +2  \bar \rho_B(x)-1) \psi_A(x) + 2 \bar\rho_A(x)\psi_B(x) \Big) \\
 \partial_x^2\psi_B(x) + \beta \partial_x \Big(  (1-2 \bar \rho_A(x) -  2 \bar \rho_B(x)) \psi_B(x) - 2\bar\rho_B(x)  \psi_A(x) \Big) \\
\end{array} \right)
\label{L_rond}
\end{equation}
It is easy to check using (\ref{eq: hydro1-bis}) that
\begin{equation}
{\cal L}  \  \bar\rho'(x)=0
\label{zero-mode}
\end{equation}
 as expected from the translation invariance \eqref{profiles}.

As the dynamics conserve the total number of particles of each species, the fluctuating parts $\phi_A(x,\tau)$ and $\phi_B(x,\tau)$ satisfy
\begin{equation}
\int \phi_A(x,\tau) dx = 
\int \phi_B(x,\tau) dx = 
0
\label{phi-phi}
\end{equation}
Let us assume that $ \bar\rho'(x)$ is the only eigenvector of ${\cal L} $  with a zero eigenvalue in the space of functions which satisfy (\ref{phi-phi}) and that all the other right eigenvectors $\psi^{(\alpha) } $ of ${\cal L}$  (for $\alpha \ge 1$) have eigenvalues $\lambda_\alpha$ with negative real part
\begin{equation}
{\cal L}   \  \psi^{(\alpha) } (x)  = \lambda_\alpha \   \psi^{(\alpha) } (x)  
\label{eigenvalue}
\end{equation}
In the following we will  also use the scalar product between a left vector $\chi(x)=( \chi_A(x),\chi_B(x))$ and a right vector $\psi(x)= 
 \left(\begin{array}{l}
\psi_A(x) \\
\psi_B(x)
\end{array} \right)
$
\begin{equation}
\label{scalar-product}
\langle \chi| \psi \rangle = \int_0^1 \Big[\chi_A(x) \psi_A(x) +  \chi_B(x) \psi_B(x) \Big] dx
\end{equation}
It is  easy to check, using (\ref{L_rond}), integrations by parts and (\ref{eq: hydro1-bis}), that    for any right vector $\psi(x)$
\begin{equation}
\label{scalar-product1}
\langle \chi^{(0)} | {\cal L} \  \psi \rangle = 0
\end{equation}
where the vector  $\chi^{(0)} (x)$ is given by
\begin{equation}
 \chi^{(0)}(x)
=(\bar\rho_B(x),- \bar\rho_A(x)) 
\label{left-zero-mode}
\end{equation}
So   (\ref{left-zero-mode})   is a left eigenvector of ${\cal L}$ with  zero eigenvalue.
It follows  that for all the eigenvectors $\psi^{(\alpha)}(x)$  with a non-zero eigenvalue $\lambda_\alpha$
one has
\begin{equation}
<\chi^{(0)} | {\cal L} \psi^{(\alpha)} > = \lambda_\alpha <\chi^{(0)} | \psi^{(\alpha)} >  = 0
\label{orthogonality}
\end{equation}

If we put the expressions  (\ref{noisy-profiles}) into the equations of evolution of the density profiles  (\ref{qaqb}, \ref{conservations}) one gets at leading order in $1/\sqrt{L}$
\begin{equation}
-(\partial_\tau \xi_\tau ) \bar{\rho}'(x) 
 +{1 \over \sqrt{L}}  \  \partial_\tau \phi(x,\tau) =   {1 \over \sqrt{L}}  \ {\cal L} \phi -  {1 \over \sqrt{L}} \partial_x \eta(x,\tau)
\label{evol1}
\end{equation}
From \eqref{evol1}, one can see already that the variance of $\xi_\tau$ scales as $1/L$ as  claimed in (\ref{main-result}). 
This $L$ dependence is consistent with the $L^3$ dependence of the relaxation time found in   \cite{BCP}  in the case of equal densities ($r_A=r_B=r_C=1/3$) (see section 6 and figure 1 below).

Under the assumption that
one can decompose $ \partial_x \eta(x,\tau)$ on the eigenvectors of ${\cal L} $
and that $\bar\rho'(x)$ is the only eigenvector with zero eigenvalue satisfying (\ref{phi-phi}),  one can write
$$
\partial_x \eta(x,\tau) = c_0(\tau) \bar{\rho}'(x)  +  \sum_{\alpha \ge 1} c_\alpha(\tau) \psi^{(\alpha)}(x)  
$$
This  together with (\ref{evol1}) leads to
$$ \phi(x,\tau) = - \sum_{\alpha  \ge 1}\psi^{(\alpha)}(x) \int_{-\infty}^\tau c_\alpha(\tau') e^{ \lambda_\alpha(\tau-\tau')}
 d \tau' $$
and to
\begin{equation}
\partial_\tau \xi_\tau  
=   { c_0(\tau) \over \sqrt{L} }
 \label{evol2}
\end{equation}
Doing so, we have imposed that $\phi$ has a zero component on $\bar \rho'(x)$ (the component on $\bar \rho'(x)$  of the r.h.s. of 
\eqref{evol1} can always be absorbed as a translation which contributes to $\xi_\tau$).

As $\phi(x)$ has only components on the right eigenvectors of ${\cal L}$ with non zero eigenvalue, 
one can determine $c_0(\tau) $  by taking the scalar product of   
(\ref{evol1})  with  the left eigenvector $\chi^{(0)}(x)$ and by using the orthogonality property (\ref{scalar-product1}, \ref{orthogonality}) one gets
\begin{equation}
\partial_\tau \xi_\tau  
={ 1 \over \sqrt{L}}{ < \chi^{(0)}(x)| \partial_x \eta(x,\tau) > \over  < \chi^{(0)}(x)| \bar\rho^\prime (x) > }
 \label{evol3}
\end{equation}
which after an integration by parts becomes 
\begin{equation}
\partial_\tau \xi_\tau  
=- { 1 \over \sqrt{L}}{ < \partial_x \chi^{(0)}(x)|  \eta(x,\tau) > \over  < \chi^{(0)}(x)|  \bar\rho^\prime (x) > }
 \label{evol4}
\end{equation}
This together with (\ref{sigma},\ref{noise},\ref{Sigma})
leads to the following expression for the diffusion constant  $D$ defined in (\ref{main-result}) 
\begin{equation}
D= {< \partial_x \chi^{(0)}(x)|  \Sigma(x) | \partial_x \chi^{(0)}(x) >  \over  < \chi^{(0)}(x)| \bar\rho^\prime (x) > ^2}
\label{D-exp}
\end{equation}
where we have used the fact (\ref{noise},\ref{Sigma})  that $\Sigma(x)$ is the covariance matrix of the noise $\eta(x,\tau)$.
From   (\ref{Sigma}, \ref{left-zero-mode}) one deduces
$$
D= {  \int dx  \big( 2 \bar \rho_A (1-  \bar\rho_A) \rho_B'^2  + 4   \bar\rho_A  \bar\rho_B  \bar\rho_A'  \bar \rho_B' + 2  \bar\rho_B (1-  \bar\rho_B)  \bar\rho_A'^2 \big) \over 
 \left[\int dx  \left(  \bar\rho_A  \bar\rho_B' -  \bar\rho_A'  \bar\rho_B\right) \right]^{2}}
$$ which can be rewritten as 
(\ref{main-result-ter}) using (\ref{conservation}).

Although (\ref{noisy-profiles}) should remain valid for times $\tau \gg L$, we have neglected in the derivative (\ref{evol1},\ref{evol4}) of  $\xi_\tau$ corrections of order $1/L$. Integrating (\ref{evol4}) gives $\xi_\tau \sim \sqrt{\tau \over L}$ which might compete with  a possible drift correction  of order ${\tau\over L}$. Therefore the value of $\xi_\tau$  obtained by integrating (\ref{evol4}) should only be valid for $\tau \ll L$.

\section{The large  deviation  approach}

In the large deviation approach we 
 compute the probability (\ref{pro}, \ref{eq: I ABC}) of observing  profiles   close to the  steady state profiles $\bar\rho_A(x),\bar\rho_B(x)$  moving for a long time at a given velocity   $v$ 
around the circle.

For small $v$ one expects these profiles to be of the form 
\begin{eqnarray}
\rho_A(x,\tau) = \bar\rho_A(x-v\tau) + v  \  \psi_A(x-v\tau) + O(v^2)
\nonumber \\
\label{moving-profiles}
\rho_B(x,\tau) = \bar\rho_B(x-v\tau) + v  \  \psi_B(x-v\tau) + O(v^2)
\end{eqnarray}
where $\psi_A$ and $\psi_B$ are the deformations of the profiles due to the velocity $v$:
we  have to look for the deformations $\psi_A,\psi_B$ which  maximize  the probability of such moving profiles.
To produce such time dependent density profiles, the currents \eqref{conservations} should be of the form
(at first order in $v$)
\begin{eqnarray}
j_A(x,t) = I_A + v  \ i_A+ v \ \bar\rho_A(x-vt) + O(v^2) \nonumber \\
j_B(x,t) = I_B  + v  \ i_B + v \ \bar\rho_B(x-vt) + O(v^2)
\label{eq: courant fluc}
\end{eqnarray}
where $I_A,I_B$  are the steady state currents, which satisfy
\begin{eqnarray}
I_A =  -\partial_x \bar\rho_A   - \beta  \bar\rho_A ( \bar\rho_A + 2 \bar\rho_B-1)
\nonumber \\
\label{eq: hydro1-ter}
I_B=  -\partial_x \bar\rho_B   - \beta   \bar\rho_B (1-2  \bar\rho_A - \bar\rho_B) 
\end{eqnarray}
$i_A,i_B$ are additional currents and the terms 
$ v \bar\rho_A(x-vt), v \bar\rho_B(x-vt)$ are there to insure that the conservation laws (\ref{conservations}) are satisfied.
For small velocity $v$, the cost of the large deviation functional (\ref{eq: I ABC}) evaluated for the moving profiles 
\eqref{moving-profiles} can be written as
\begin{eqnarray}
\cC & = & v^2 \int_0^1 dx \
\frac{\bar\rho_B (1-\bar\rho_B)  (i_A +\bar\rho_A - \phi_A)^2}{4 \bar\rho_A \bar\rho_B (1-\bar\rho_A - \bar\rho_B)}  
+ \frac{\bar\rho_A(1-\bar\rho_A) (i_B +\bar\rho_B - \phi_B)^2 }{4 \bar\rho_A \bar\rho_B (1-\bar\rho_A - \bar\rho_B)}  \nonumber \\
&& \qquad \qquad + 2 \frac{ \bar\rho_A \bar\rho_B  (i_A +\bar\rho_A - \phi_A) (i_B +\bar\rho_B - \phi_B)}{4 \bar\rho_A \bar\rho_B (1-\bar\rho_A - \bar\rho_B)}  
\label{cost}
\end{eqnarray}
where
\begin{eqnarray}
\label{M_rond}
&& \phi(x) = {\cal M} \psi(x) \\
&& \quad = \left(\begin{array}{l}  -\partial_x \psi_A(x) -
\beta  \Big((2 \bar\rho_A(x) + 2 \bar \rho_B(x)-1) \psi_A(x)  + 2 \bar\rho_A(x) \psi_B(x) \Big) \\
 -\partial_x \psi_B(x) -
\beta  \Big((1-2 \bar\rho_A(x) - 2 \bar \rho_B(x)) \psi_B(x)  - 2 \bar\rho_B(x) \psi_A(x) \Big) 
\end{array} \right) \nonumber
\end{eqnarray}
is obtained by linearizing (\ref{qaqb}).
One can notice from (\ref{L_rond}) that for any right vector $\psi(x)$ 
\begin{equation}
\partial_x ( {\cal M} \psi(x)) = - {\cal L} \psi(x) 
\label{M_L}
\end{equation}
This implies that (see (\ref{left-zero-mode},\ref{orthogonality})) that $\partial_x \chi^{(0)}(x) =(\bar\rho_B'(x),-\bar\rho_A'(x))$ is a left eigenvector of ${\cal M}$ with a zero eigenvalue and therefore that for any right vector $\psi(x)$, one has 
\begin{equation}
  < \partial_x \chi^{(0)}(x) | {\cal M} | \psi(x) > = 0 
\label{orthogonality-bis}
\end{equation}

Using the scalar product (\ref{scalar-product}) one can rewrite  (\ref{cost}) as 
\begin{equation}
\cC = 
{v^2  \over 2} < i+\bar\rho(x)-\phi(x) \ | \Sigma^{-1}(\bar \rho(x))  | \  i+\bar\rho(x)-\phi(x) >
\label{cost-1}
\end{equation}
 where the inverse of the matrix $\Sigma( \bar \rho(x))$ defined in (\ref{Sigma}) is 
\begin{equation}
\label{Sigma-1}
\Sigma^{-1}( \bar \rho(x)) = {1 \over 2 \bar\rho_A(x) \bar\rho_B(x) \bar \rho_C(x) } \left(\begin{array}{cc}   \bar\rho_B(x)(1- \bar\rho_B(x)) &  \bar\rho_A(x)\bar\rho_B(x)
\\\bar\rho_A(x)\bar\rho_B(x)  &  \bar\rho_A(x)(1- \bar\rho_A(x)) \end{array} \right) 
\end{equation}
with $\bar \rho_C(x) = 1- \bar\rho_A(x)-\bar\rho_B(x)$.
We have now to optimize  this expression over the currents $i$ and the deformations $\psi$ which appear in (\ref{cost-1}) through (\ref{M_rond}).
 To do so let us consider the right vector $\partial_x( \Sigma( \bar\rho(x) ) \partial_x \chi^{(0)}(x)) $. As it is a derivative, this vector has a zero average  
(\ref{phi-phi}) on the circle and  we assume that it can be decomposed on the basis of the eigenvectors of ${\cal L}$
$$\partial_x( \Sigma( \bar\rho(x) ) \partial_x \chi^{(0)}(x))  = b_0 \bar\rho'(x) + \sum_{\alpha \ge 1} b_\alpha \psi_\alpha(x) = b_0 \bar\rho'(x) + {\cal L} \hat\psi(x)$$
where $\hat\psi(x)= \sum_\alpha \frac{b_\alpha}{\lambda_\alpha} \; \psi_\alpha(x)$
or equivalently  
\begin{equation}
\label{eq: decomposition vecteur}
\bar\rho'(x) = a \partial_x( \Sigma(\bar \rho(x)) \partial_x \chi^{(0)}(x))  + {\cal L} \tilde\psi(x)
\end{equation}
(with $a=1/b_0$ and $\tilde\psi=-\hat\psi/b_0$). 
Integrating over $x$ using (\ref{M_L}) one gets
\begin{equation}
\bar\rho(x) = a  \Sigma(\bar \rho(x)) \partial_x \chi^{(0)}(x)  - {\cal M} \tilde\psi(x) + \tilde{i}
\label{decomposition_de_rho}
\end{equation}
where $\tilde{i}$ is just the constant of integration.
Then replacing $\bar\rho(x)$ by  (\ref{decomposition_de_rho}) into
(\ref{cost-1}),   one gets using the orthogonality property   (\ref{orthogonality-bis})  
\begin{eqnarray}
\cC = 
{v^2 \over 2} \left[< i + \tilde{i} - {\cal M}( \psi+\tilde\psi)  | \Sigma^{-1}(\bar \rho(x))   | i + \tilde{i} - {\cal M}( \psi+\tilde\psi) >
\right. \nonumber \\ \left.
 +a^2 <  \partial_x \chi^{(0)}(x) | \Sigma(\bar \rho(x)) | \partial_x \chi^{(0)}(x) > \right]
\label{cost-2}
\end{eqnarray}
As the matrix $\Sigma^{-1}$  (see (\ref{Sigma-1})) is positive definite, it is easy to see that the optimal choice for $i$ and 
$\psi$ is $i=-\tilde{i}$ and $\psi= -\tilde{\psi}$.   By multiplying (\ref{decomposition_de_rho}) by the left vector 
$\partial_x \chi^{(0)}(x) $ and by using the orthogonality property (\ref{orthogonality-bis}), one gets for the projection $a$ in (\ref{decomposition_de_rho})
$$ a ={ < \partial_x \chi^{(0)}(x) | \bar \rho(x)  > \over   < \partial_x \chi^{(0)}(x)|  \Sigma(\bar \rho(x)) |\partial_x \chi^{(0)}(x)>}
$$
Therefore replacing $i,\psi, a$ by their expressions into (\ref{cost-2}) gives for the probability of seeing  moving steady state profiles at a small velocity $v$
during some  macroscopic time $\tau$
\begin{eqnarray}
\label{eq: deviation vitesse}
{\rm Pro}(v)  \sim  
\exp \left( - L \min_{\psi, i} \cC \right) 
= \exp \left[- L {v^2 \over 2}  \tau  { < \partial_x \chi^{(0)}(x) | \bar \rho(x)  >^2 \over   < \partial_x \chi^{(0)}(x)|  \Sigma(\bar \rho(x)) |\partial_x \chi^{(0)}(x)>}
\right] 
\end{eqnarray}

In appendix II, we give another derivation of \eqref{eq: deviation vitesse} in the reversible case $r_A =r_B = r_C = 1/3$.
\begin{rem}
\label{rem H-1}
Instead of the scalar product \eqref{scalar-product}, it could be also convenient (see for example \cite{BGP}) 
to consider the modified scalar product
\begin{eqnarray}
\label{eq: norm -1}
\bra f \, \big| \, g \ket_{-1,\cond}
= \left \bra \nabla^{-1} f  \, \big| \,  \cond^{-1} (\bar \rho)  \, \big| \, \nabla^{-1} g \right \ket  
\end{eqnarray}
where $ \nabla^{-1} f$ is defined for vectors $f (x)= \binom{f_A(x)}{f_B(x)}$ with zero mean  \eqref{phi-phi} by
\begin{eqnarray}
\label{eq: constant -1}
\nabla^{-1} f (x) = \int_0^x du f(u) + K, 
\end{eqnarray}
and the constant $K$ is fixed such that $\left \bra  \cond(\bar \rho)^{-1}   \; \nabla^{-1} f \right \ket  = 0$. 

The large deviation cost \eqref{cost-1} can be rewritten
\begin{equation*}
\cC = \frac{v^2}{2}  \big \bra \partial_x \bar\rho- \partial_x \phi \, \big| \, \partial_x \bar\rho- \partial_x \phi \big \ket_{-1,\cond}
\end{equation*}
The optimal value of $i$ in \eqref{cost-1} is implicitly chosen by the condition \eqref{eq: constant -1}. 
From the identity \eqref{M_L}, one finally gets
\begin{equation*}
\cC = \frac{v^2}{2}  \big \bra  \bar\rho'- {\cal L} \psi \, \big| \, \bar\rho'-  {\cal L} \psi \big \ket_{-1,\cond}
\end{equation*}
The latter expression provides another way to understand the decomposition \eqref{eq: decomposition vecteur}.
\end{rem}

The large deviation result \eqref{eq: deviation vitesse} was obtained in the limit where $\tau$ and $L$ go to $\infty$  first and then $v$ tends to 0.
If we rewrite \eqref{eq: deviation vitesse} in terms of the displacement $\xi_\tau= v \tau$, we get
\begin{eqnarray}
\label{eq: proba xi}
{\rm Pro}(\xi_\tau)  \sim  \exp \left[- L {\xi_\tau^2 \over 2 \tau}    { < \partial_x \chi^{(0)}(x) | \bar \rho(x)  >^2 \over   < \partial_x \chi^{(0)}(x)|  \Sigma(\bar \rho(x)) |\partial_x \chi^{(0)}(x)>} \right]
\end{eqnarray}

Formally, this leads to the same expression for the diffusion constant (\ref{main-result}) as (\ref{D-exp}). 
The difficult to recover rigorously the diffusion coefficient is that   (\ref{eq: proba xi})   has been obtained 
 for  $\xi_\tau $  of order  $  \tau$ while to extract the diffusion constant from (\ref{eq: proba xi})  one would need  $\xi_\tau$ to be of order 
$\sqrt{\tau \over L}$, that is for $v$ of order $\sqrt{1 \over \tau  L}$.  The fact that  (\ref{eq: proba xi}) does give  the  right diffusion constant means that some kind of exchange of the order of limits $\tau,L \to \infty$ and $v \to 0$ should hold.

As we will see in the simulations of section \ref{sec: simulations} (Figure 2), $\xi_\tau$ may have a drift of order $\tau/L$ which is not predicted by 
\eqref{eq: proba xi}.

\section{Two particular cases}

We now briefly discuss two cases where the expression (\ref{main-result-ter}) simplifies.

\subsection{The equal density case}

When the densities of the 3 species are equal ($r_A=r_B=r_C=1/3$), the dynamics satisfy detailed balance.
Therefore the currents $I_A$ and $I_B$ in (\ref{eq: hydro1-ter}) vanish and the steady state profiles satisfy 
\begin{eqnarray}
\partial_x \bar\rho_A  =  \beta  \big( \bar\rho_A ( \bar\rho_C - \bar\rho_B )  \big)
\nonumber  \\
\partial_x \bar\rho_B  =  \beta  \big( \bar\rho_B ( \bar\rho_A - \bar\rho_C ) \big) 
\label{prof} \\
\partial_x \bar\rho_C  =  \beta  \big( \bar\rho_C ( \bar\rho_B - \bar\rho_A ) \big) 
\nonumber
\end{eqnarray}
This implies that
  $\partial_x (\bar\rho_A  \bar\rho_B  \bar\rho_C ) =0 $  and
\begin{equation}
\bar\rho_A(x) \    \bar\rho_B(x)  \  \bar\rho_C(x)  = \Gamma
\label{product}
\end{equation}
where $\Gamma$ is a constant  as already noted in \cite{ACLMDS}.
\\
By replacing $\bar\rho_C'(x)^2 $ by $\beta \bar\rho_C'(x) \bar\rho_C(x)(\bar\rho_B(x)-\bar\rho_A(x))$ and doing  similar changes for $\bar\rho_A'(x)^2 $  and $\bar\rho_B'(x)^2 $ in the numerator of (\ref{main-result-ter}) one gets 
\begin{align*}
& 2  \int dx  \left(  \bar \rho_A   \bar\rho_B \bar\rho_C'^2  +    \bar\rho_B  \bar\rho_C  \bar\rho_A'^2 +   \bar\rho_C  \bar\rho_A  \bar\rho_B'^2 \right) 
\\
& \quad = 2 \beta \Gamma  \int dx [
\bar\rho_C'(x) (\bar\rho_B(x)-\bar\rho_A(x)) + 
\bar\rho_A'(x) (\bar\rho_C(x)-\bar\rho_B(x)) + 
\bar\rho_B'(x) (\bar\rho_A(x)-\bar\rho_C(x)) ] 
\\
& \quad =  6  \beta \Gamma  \int dx [
\bar\rho_A(x) \bar\rho_B'(x)-
\bar\rho_B(x) \bar\rho_A'(x) ]
\end{align*}
Therefore the ratio (\ref{main-result-ter}) becomes
$$ D={6 \beta \Gamma  \over  \int dx [
\bar\rho_A(x) \bar\rho_B'(x)-
\bar\rho_B(x) \bar\rho_A'(x) ]
} 
$$
one can further transform the denominator  using  (\ref{prof})
\begin{eqnarray*}
\int dx [ \bar\rho_A \bar\rho_B'- \bar\rho_B \bar\rho_A' ]
= \beta 
\int dx [ \bar\rho_A \bar\rho_B(1- 3 \bar\rho_C ) ] = \beta \Gamma \left[ \int {dx \over  \bar\rho_C(x)} -3 \right]
\end{eqnarray*}

Finally  the expression (\ref{main-result-ter})  becomes
\begin{equation}
\label{D-reversible}
D= 6  \left[    \int  {dx \over \bar \rho_C(x)} -3  \right]^{-1}
\end{equation}
(in the equal density case \eqref{D-reversible} would of course remain unchanged by replacing $\bar \rho_C$ by $\bar \rho_A$ or $\bar \rho_B$).

\subsection{Close to the second order phase transition}

When the phase transition of the ABC model on the ring is second order, and $\beta$ approaches $\beta_c$ from above,  the steady state profiles become sine functions \cite{CDE}

\begin{eqnarray*}
 \rho_A(x) &\simeq&  r_A+ \Psi  \left[\sqrt{r_A}  \  e^{2 i \pi (x-x_0)} +
c.c. \right]
\label{rhoa}\\[1ex]
 \rho_B(x) &\simeq& r_B+ \Psi  \left[ {r_C - r_A - r_B - i \sqrt{ 1 - 2
(r_A^2 + r_B^2 + r_C^2) } \over 2 \sqrt{r_A}}   e^{2 i \pi
(x-x_0)} +
c.c. \right]
\label{rhob}\\[1ex]
\rho_C(x) &\simeq& r_C + \Psi  \left[ {r_B - r_A - r_C + i \sqrt{ 1 - 2
(r_A^2 + r_B^2 + r_C^2) } \over 2 \sqrt{r_A}}   e^{2 i \pi
(x-x_0)} +
c.c. \right]
\label{rhoc}
\end{eqnarray*}
where 
\begin{equation}
\Psi =  
 {1 - 2
(r_A^2+r_B^2+r_C^2) \over \sqrt{2 (r_A^2 +
r_B^2 + r_C^2) - 4 (r_A^3 + r_B^3 + r_C^3) }}
\ \left(   \frac{\beta -\beta_c}{\beta_c}\right) ^{1/2}
\label{op}
\end{equation}
Then the integrals which appear in (\ref{main-result-ter}) become for $\beta$ close to $\beta_c$
$$ 2  \int dx  \left(  \bar \rho_A   \bar\rho_B \bar\rho_C'^2  +    \bar\rho_B  \bar\rho_C  \bar\rho_A'^2 +   \bar\rho_C  \bar\rho_A  \bar\rho_B'^2 \right)
 \simeq 48   \pi^2 r_A r_B r_C  \ \Psi^2$$
and 
$$  \int dx  [ \bar\rho_A \bar\rho_B' - \bar\rho_B \bar\rho_A'] 
\simeq  4  \pi \sqrt{ 1 - 2 r_A^2 - 2 r_B^2 - 2 r_C^2} \  \Psi^2$$

So that the diffusion constant $D$ diverges as $(\beta-\beta_c)^{-1}$
\begin{eqnarray*}
D  &\simeq&  { 3 r_A r_B r_C \over  (1 - 2 r_A^2 - 2 r_B^2 - 2 r_C^2) \Psi^2} \\
&& \qquad =  { 6 r_A r_B r_C \big(r_A^2 +
r_B^2 + r_C^2 - 2 (r_A^3 + r_B^3 + r_C^3)\big)   \over (1- 2
(r_A^2+r_B^2+r_C^2))^3 } \ {\beta_c \over \beta-\beta_c} 
\end{eqnarray*}

\section{Simulations}
\label{sec: simulations}

From (\ref{eq: 4 bis}) 
one can deduce the long time decay of the autocorrelation function: if $S_k(t)=A, B $ or $C$ is the type of particle on site $k$ at time $t $, then in the steady state, for times $|t-t'|= L^2 \tau \gg L^2$
\begin{eqnarray}
&&  {\rm Prob}( S_k(t) =  S_k(t') ) =
\\
&&  \ \ \ \ 
\mathbbm{E} _{\xi_\tau}
\left[\int dx \bar\rho_A(x) \bar\rho_A(x-\xi_\tau) + \bar\rho_B(x) \bar\rho_B(x-\xi_\tau)+\bar\rho_C(x) \bar\rho_C(x-\xi_\tau)  \right]
\nonumber
\end{eqnarray}
where the expectation $\mathbbm{E} _{\xi_\tau}$ is  over the Gaussian variable $\xi_\tau$.
Assuming that $\xi_\tau$ has no drift and  behaves as a Brownian motion with diffusion coefficient 
(\ref{main-result}, \ref{main-result-ter}) then one gets in microscopic units (\ref{scaling})
\begin{eqnarray}
&& {\rm Prob}( S_k(t) =  S_k(t') ) -  (r_A^2 + r_B^2 + r_C^2)  \nonumber \\
&& \qquad \qquad = \sum_{n \neq 0}   (|a_n|^2 + |b_n|^2 + |c_n|^2) e^{-2 n^2 \pi^2  D \tau / L}
\nonumber \\
&& \qquad  \qquad =  \sum_{n \neq 0}   (|a_n|^2 + |b_n|^2 + |c_n|^2) e^{-2 n^2 \pi^2  D |t'-t| / L^3} 
\label{exponential}
\end{eqnarray}
In (\ref{exponential})  the $a_n$ are the Fourier coefficients of the  steady state density  profiles
\begin{equation}
\bar \rho_A(x) = r_A + \sum_{n \neq 0} a_n e^{2 i \pi n x}
\label{fourier}
\end{equation} 
(with similar definitions for $b_n$ and $c_n$)  and  $r_A, r_B, r_C$ are the global densities of the three species.
Should $\xi_\tau$ have a drift of order $1/L$ (as measured in the non equilibrium case), then additional oscillations would occur in the autocorrelation function \eqref{exponential}.

In the equal  density case ($r_A=r_B=r_C=1/3$),  we have performed Monte Carlo simulations of the ABC model  to measure the autocorrelation function  predicted by \eqref{exponential}. 
At each time step we also measured the first Fourier component of the species $A$
\begin{equation}
\label{eq: phase}
U(t) e^{i V(t)}= \sum_{k =1}^L e^{2 i \pi k /L} \, 1_{ \{  S_k(t) =A \}}
\end{equation}
 and the autocorrelation  $\langle \cos(V(t') - V(t)) \rangle$ of the phase $V(t)$. 
The phase $V(t)$ is related to the displacement $\xi_\tau$ by $V ( \tau L^2) \equiv 2 \pi \,  \xi_\tau$ where the equality is modulo $2 \pi$.
Our results  are shown in a semilogarithmic scale in Figure 1 for three different system sizes $L=60, 120, 240$ when $\beta=15$.
The data for the three different sizes overlap, indicating that the right time scale is $L^3$.
There is also  an excellent agreement with the prediction (\ref{main-result-ter}), which in this case would give $D \simeq .94$ estimated by an independent calculation of the profiles obtained by looking at the solution of the deterministic equations (\ref{eq: hydro1-bis}).

\begin{figure}[ht]
\centerline{
\includegraphics[width=8cm]{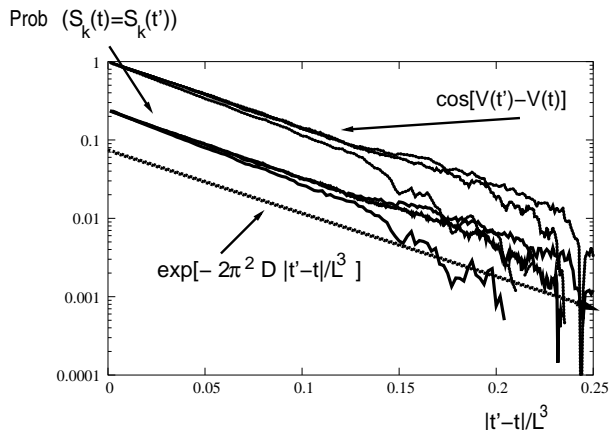}
}
\caption{\small The exponential decay of the autocorrelation   ${\rm Prob}( S_k(t) =  S_k(t'))$ (lower curves)  and of $\langle \cos(V(t') - V(t)) \rangle$  versus $|t-t'|/L^3$ in the equal density case, for $\beta=15$, and for 3 system sizes: $L=60,120, 240$. The thin line indicates the slope predicted by our main  result (\ref{main-result-ter}).}
\end{figure}

\medskip

In attempts to check that the steady state profiles have no velocity, as claimed in the appendix I, we
performed simulations of the ABC model for 3 system sizes ($L=60, 120, 240$) at densities $r_A=1/3, r_B=1/2, r_C=1/6$
and $\beta = 17$.
The position  $\xi_\tau $ of the profiles was measured  by following the phase $V(t) \equiv 2 \pi \xi_\tau$ of the first
Fourier mode of the density profile of the species $A$ \eqref{eq: phase}. 
Our results in Figure 2 (left) show that for these densities  
the displacement $\xi_\tau$  has a linear time dependence and therefore the profiles have
a non-zero velocity, but this velocity decreases with the system size. When we multiply the displacement by  the system size $L$ as in Figure 2 (right), we see that the three sets of data overlap
indicating that the velocity scales as $1/L$.
We observed this $1/L$  decay of the velocity for other choices of the densities, but the decay looked slower when we approached the phase transition line.
We did not succeed in computing this velocity $v_L(r_A,r_B,r_C)$ but we think that it should be an antisymmetric function of the three densities and therefore should vanish   whenever two densities are equal.
Understanding this velocity of order $1/L$ would require to rewrite the fluctuating hydrodynamics equations \eqref{evol1} to a higher order in $1/L$ or to calculate corrections of order 1 in the large deviations \eqref{pro}.
For Ginzburg-Landau equation with white noise, it has been shown in \cite{BB} that the interplay between the non-linearity and the noise may induce  such a small drift. It would be interesting to see if an analogous approach could be used for  the ABC model.

\begin{figure}
\centerline{\includegraphics[width=6cm]{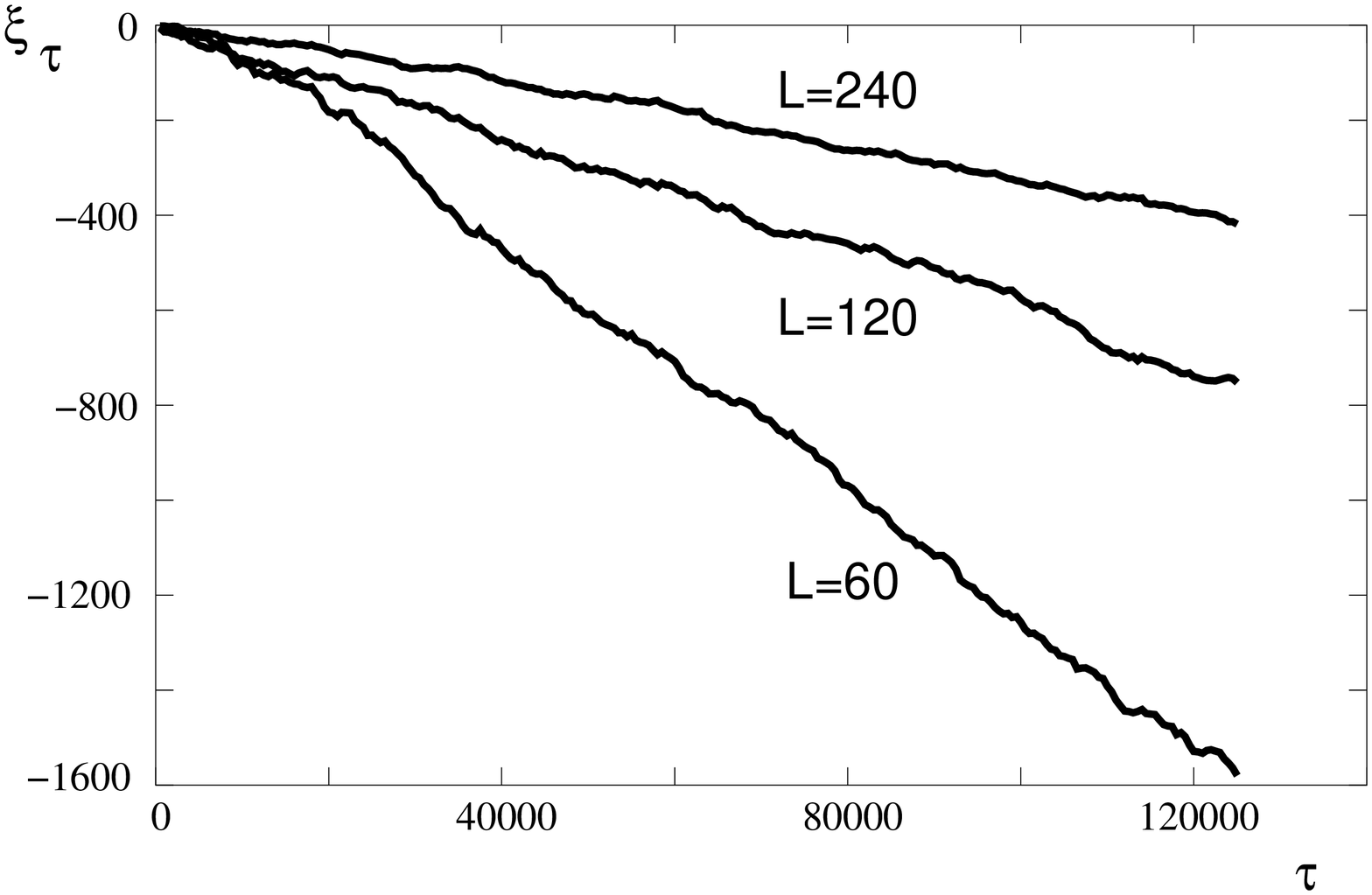}
\hskip1cm
\includegraphics[width=6cm]{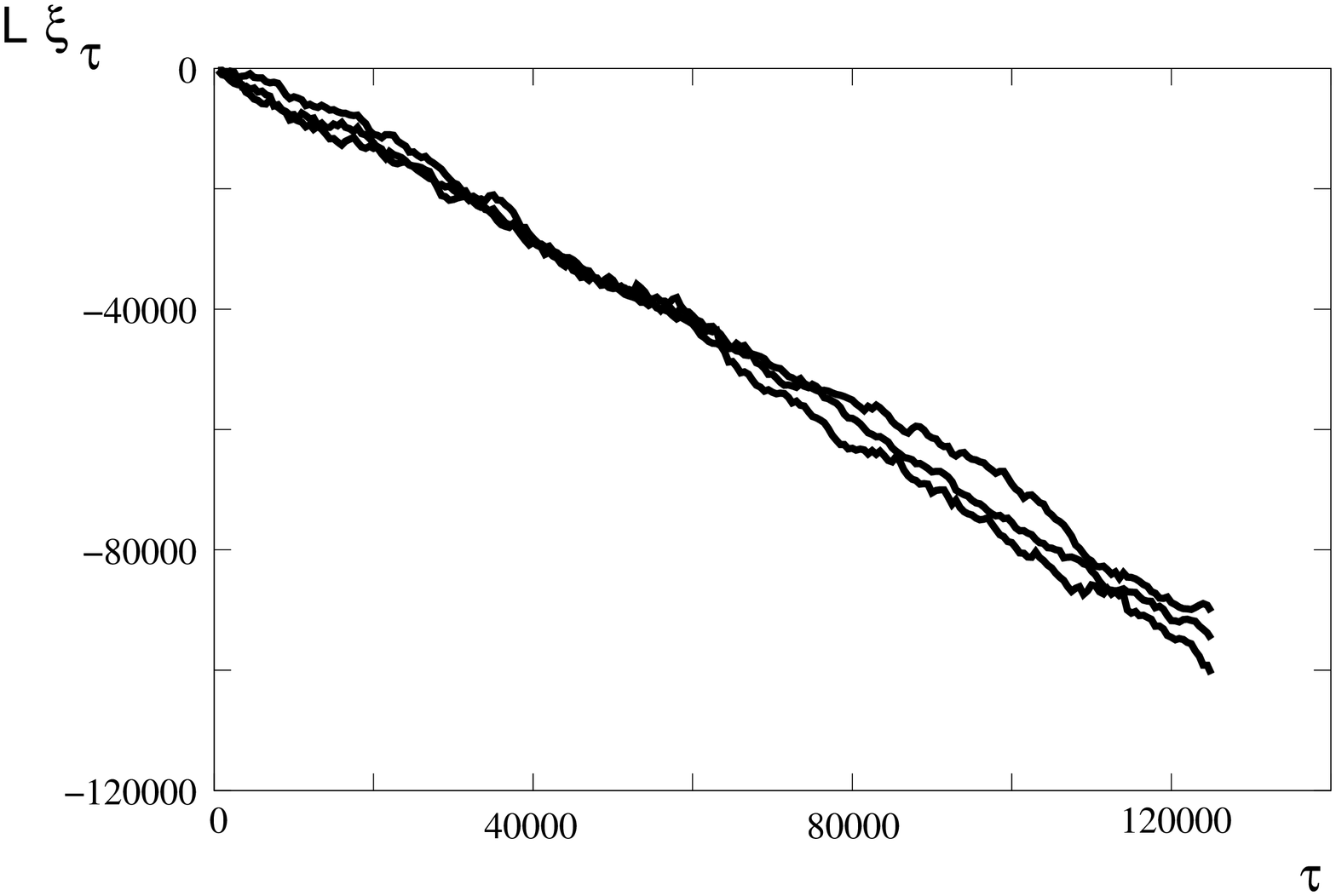}
}
\caption{\small On the left, the linear displacement of $\xi_\tau$ for three system sizes $L=60,120, 240$ at densities 
$r_A=1/3, r_B=1/2, r_C=1/6$ and $\beta = 17$. 
On the right, the same data are depicted after rescaling by $1/L$.  }
\end{figure}

\section{Conclusion}

In this paper, we have studied the fluctuations of the steady state profiles of the ABC model in the  segregated phase. 
We have shown that the position of the density profiles fluctuates on a microscopic time scale of order $L^3$ and its variance has been computed in terms of the steady state profiles \eqref{main-result-ter}. 
The result has been obtained by using fluctuating hydrodynamics and a large deviation approach. 
It has been confirmed by numerical simulations (see section \ref{sec: simulations}).

Our result relies on the assumptions that the solutions of the macroscopic  steady state equations \eqref{eq: hydro1} are unique (up to translations) and do not move. This claim is correct at equal densities but it would be important to determine the range of parameters for which it remains valid.  Both methods we used (linearized fluctuating hydrodynamics and large deviations) are valid for macroscopic time scales and as the fluctuations of the steady state occur on a longer time scale we had to extrapolate the results given by these methods. Several questions remain in order to understand the validity of this extrapolation as well as the intriguing small drift of order $1/L$ observed in the numerical simulations (see figure 2).

\vskip1cm
It is our pleasure to  dedicate this work to Professor Cyril Domb on the occasion of his 90th birthday.
\vskip1cm

\noindent
{\it Acknowledgments.}
We thank A. Gerschenfeld, G. Giacomin and D. Mukamel for very helpful discussions.
TB and BD acknowledge the support of the French Ministry of Education through the ANR 2010 BLAN 0108 01 grant. 
We thank one of the referees for pointing out the reference \cite{BB}.

\section*{Appendix I}

In this appendix, we argue that for densities $r_A, r_B, r_C$ close to the equal density case, the steady state profile solution of  
\eqref{eq: hydro1} do not move.

Let us assume that the steady state profiles of the macroscopic equations \eqref{eq: hydro1} have the form 
$\bar \rho (x - v \tau) = \{ \bar \rho_A  (x - v \tau), \bar \rho_B  (x - v \tau) \}$.
When $r_A= r_B = 1/3$, the dynamics are reversible.
The invariant measure is known and the steady state profiles have no velocity and are solutions of \eqref{eq: hydro1-bis}.
We are going to check that this property holds for general mean densities provided smoothness assumptions are  satisfied. 

Suppose that for the mean densities $\{r_A, r_B\}$, the steady state profile $\bar \rho = \binom{\bar \rho_A}{ \bar \rho_B}$ has a velocity $v$ and is solution of \eqref{eq: hydro1}. For a small shift of the densities $\{ r_A + \delta_A, r_B + \delta_B \}$, we assume that the steady state profile $\hat \rho = \binom{\hat \rho_A}{ \hat \rho_B}$ is obtained as a perturbation of $\bar \rho$
\begin{eqnarray}
\label{eq: perturbation densite}
\begin{cases}
\hat \rho_A (x,\tau) = \bar \rho_A (x - v_\delta \, \tau ) + \psi_A (x - v_\delta \, \tau )  \\
\hat \rho_B (x,\tau)= \bar \rho_B (x - v_\delta \, \tau ) + \psi_B (x - v_\delta \, \tau )  
\end{cases}
\end{eqnarray}
where $v_\delta - v$ and $\psi =  \binom{\psi_A}{\psi_B}$ are of order $\delta$.
As $\hat \rho$  is also a solution of \eqref{eq: hydro1}, one gets at the first order in $\delta$
\begin{eqnarray}
- ( v_\delta - v)  \bar \rho^\prime  - v \psi' = {\cal L} \psi 
\end{eqnarray}
where the linearized operator ${\cal L}$ has been introduced in \eqref{L_rond}.
By construction $\bar \rho^\prime$ is a right eigenvector of  $\tilde {\cal L} = {\cal L} + v \partial_x$ with zero eigenvalue.
Making the assumption that there is a corresponding left eigenvector $\tilde \chi^{(0)}$ of $\tilde {\cal L}$ with zero eigenvalue.
Then 
\begin{equation*}
( v_\delta - v) <\tilde \chi^{(0)} |  \bar \rho^\prime > =  0 
\end{equation*}
Thus the derivative of the velocity with respect to $\delta$ is zero since $ <\tilde \chi^{(0)} |  \bar \rho^\prime > \not =  0 $.
If one can interpolate by a smooth path $\{ r_A(s), r_B(s) \}_{0 \leq s \leq 1}$ from $\{1/3,1/3\}$ to $\{ r_A, r_B \}$ and 
iterate the same argument along this path, then the velocity of the steady state remains equal to zero.
This argument relies on the assumption that the profiles and the linearized operator behave smoothly with respect to small density shifts.

Let us finally add that we solved numerically the equations \eqref{eq: hydro1} for several choices of the global densities $r_A, r_B$ and in all cases the solutions converged to non moving steady states.

\section*{Appendix II}

For reversible dynamics ($r_A= r_B = 1/3$), the dynamical large deviations \eqref{pro} can be related to 
the steady state large deviations which are known explicitly \cite{CDE}.
In the steady state, the probability  of observing macroscopic density profiles
$\rho_A(x),\rho_B(x)$  has the following large $L$ dependence
\begin{equation}
 {\rm Pro} \big( \rho_A(x),\rho_B(x) \big) \sim \exp \Big( - L {\cF}[\rho_A(x), \rho_B(x) ] \Big) 
\label{F-def}
\end{equation}
with 
\begin{eqnarray}
\label{ld1}
&& {\cF}[\rho_A(x), \rho_B(x)] = \\  
&& \quad \kappa +  \int_{0}^{1} dx \left[ \rho_A(x) \ln \rho_A(x) +\rho_B(x) \ln \rho_B(x)
+\rho_C(x) \ln \rho_C(x) \right] \hspace{0.5cm} \nonumber\\
&& \quad +    \beta \int_{0}^{1} dx \int_{0}^{1}  dz \ z \left[ \rho_B(x) \rho_C(x+z) +
\rho_C(x) \rho_A(x+z) + \rho_A(x) \rho_B(x+z)\right]  \nonumber 
\end{eqnarray}
where $\rho_C(x) = 1 - \rho_A(x) - \rho_B(x)$ and  $\kappa$ is a normalisation constant.
As noted in the appendix of \cite{BDLV}, one can rewrite the evolution \eqref{eq: hydro1} of the density 
$\rho = \binom{\rho_A}{\rho_B}$ in terms of the large deviation functional $\cF$
\begin{eqnarray}
\label{eq: evolution}
\partial_\tau \rho(x,\tau) =  \partial_x \Big( \cond \big(\rho(x,\tau) \big) \partial_x   \bbF (x,\tau) \Big) 
\end{eqnarray}
where $\bbF = \binom{\bbF_A}{\bbF_B}$ is the vector defined by  
\begin{eqnarray*}
\bbF_A  (x,\tau) = {\delta {\cF} \over \delta \rho_A} \big(\rho(x,\tau) \big) , \qquad  \bbF_B  (x,\tau) = {\delta {\cF} \over \delta \rho_B} \big(\rho(x,\tau) \big) 
\end{eqnarray*}
and $\cond (\rho)$ was introduced in \eqref{Sigma}.
In particular the typical currents $q = \binom{q_A}{q_B}$ defined in \eqref{qaqb}  satisfy
\begin{eqnarray}
\label{eq: evolution current}
q(x,\tau) =  -  {1 \over 2} \cond \big(\rho(x,\tau) \big) \bbF (x,\tau) 
\end{eqnarray}
We remark that for $\rho = \bar \rho$ then $\bbF=0$ and $q =0$.

The previous identities allow us to rewrite the functional \eqref{eq: I ABC} for the joint deviation of the density $\rho =   \binom{\rho_A}{\rho_B}$
and the current $j =   \binom{j_A}{j_B}$ as
\begin{eqnarray*}
{\cI}_{[0,T]} (\rho, j) 
= \frac{1}{2}  \int_0^T d \tau   \left<  j -  q \; \Big| \;   \cond^{-1} (\rho)  \; \Big| \;  j -  q \right> 
\end{eqnarray*}
Expanding the functional one obtains
\begin{eqnarray*}
{\cI}_{[0,T]} (\rho, j) 
=\frac{1}{2}  \int_0^T d \tau     \left<  j \; \big| \; \cond^{-1} (\rho) \; \big| \;  j \right> 
+    \left< q \; \big| \; \cond^{-1} (\rho)  \; \big| \;  q \right>  
-  2    \left<  q \; \big| \; \cond^{-1} (\rho) \; \big| \;   j \right> 
\end{eqnarray*}
The cross product simplifies by using \eqref{eq: evolution current} and $\partial_t \rho = - \partial_x j$
\begin{eqnarray*}
&& 2 \int_0^T d \tau    \left< q \; \big| \;  \cond^{-1} (\rho) \; \big| \;  j \right> 
= -  \int_0^T d \tau  \left<   \cond \big(\rho \big) \partial_x  \bbF  \; \big| \;  \cond^{-1} (\rho) \; \big| \;  j \right>  \\
&& \qquad 
= -   \int_0^T d \tau  \left<  \partial_x  \bbF   \; \big| \;  j \right> 
= -  \int_0^T d \tau    \left<   \frac{\partial \cF}{\partial \rho}   \; \big| \;  \partial_t \rho \right> 
= - \cF (\rho_T) + \cF (\rho_0) 
\end{eqnarray*}
Thus one has
\begin{eqnarray}
\label{eq: I ABC 2}
{\cI}_{[0,T]} (\rho, j) = \frac{1}{2} \big( \cF ( \rho_T) - \cF (\rho_0) \big)
+  \frac{1}{2} \; \int_0^T d \tau     \left<  j \; \big| \; \cond^{-1} (\rho) \; \big| \;  j \right> 
+    \left< q \; \big| \; \cond^{-1} (\rho) \; \big| \; q \right> \nonumber \\
\end{eqnarray}

\medskip

The expression \eqref{eq: I ABC 2} of the large deviation functional shows that for small perturbations around the steady state profiles, the contributions of the current deviations and the profile deformations decouple. 
Observing a travelling wave  $\rho( x - v \tau)$ of the form \eqref{moving-profiles}
imposes a current $j (x,\tau ) = v \bar \rho( x - v \tau) + v i$ at the first order in $v$ (see \eqref{eq: courant fluc}). 
When $T$ diverges, \eqref{eq: I ABC 2} becomes at the second order in $v$ 
\begin{eqnarray*}
\cC = 
\lim_{T \to \infty} \; \frac{1}{T} {\cI}_{[0,T]} (\rho, j) =
\frac{v^2}{2}  \left \bra  \bar \rho +i  \; \big| \; \cond(\bar \rho)^{-1}   \; \big| \;   \bar \rho + i  \right \ket 
+ \left \bra  q  \; \big| \;  \cond(\bar \rho)^{-1}  \; \big| \; q \right \ket   
\end{eqnarray*}
where $q$ was defined in \eqref{eq: evolution current} in terms of $\bar \rho$. As $\cond(\bar \rho)$ is a non negative quadratic form, we see that the lowest cost is achieved when the profile $\bar \rho$ is not modified, i.e. when $\psi = 0$.
Thus it remains to optimize over the constant $i$ and the large deviation cost at the second order in $v$ is given by
\begin{eqnarray}
\label{eq: 1/diff coeff}
\cC = \lim_{T \to \infty} \; \frac{1}{T} {\cI}_{[0,T]} (\rho, j) =
\frac{v^2}{2}  \left \bra  \bar \rho + i  \; \big| \;  \cond(\bar \rho)^{-1}  \; \big| \; \bar \rho + i \right \ket  
\end{eqnarray}
with 
\begin{eqnarray}
\label{eq: normalisation i}
 i = - \left \bra  \cond(\bar \rho)^{-1}    \right \ket^{-1} 
\left \bra  \cond(\bar \rho)^{-1}   \;  \bar \rho \right \ket  
\end{eqnarray}
One can easily check that \eqref{eq: 1/diff coeff} coincides with the expression obtained in
\eqref{eq: deviation vitesse} when $r_A =r_B = r_C = 1/3$.

\end{document}